\documentclass[aps,prb,reprint,superscriptaddress]{revtex4-2}
\usepackage{silence}
\ExplSyntaxOn
\ExplSyntaxOff
\usepackage{graphicx,epstopdf,siunitx,xspace,amssymb,dsfont,bm,amsmath,float}\usepackage{braket}
\usepackage{chngcntr}
\usepackage{physics}
\usepackage[hidelinks]{hyperref}
\usepackage[capitalise]{cleveref} 
\usepackage[T1]{fontenc}
\usepackage{comment}
\usepackage{csquotes}
\usepackage[section]{placeins}
\newcommand{\ld}[0]{\ensuremath{\ket{v_0, \downarrow}}\xspace}
\newcommand{\lu}[0]{\ensuremath{\ket{v_0, \uparrow}}\xspace}
\newcommand{\ud}[0]{\ensuremath{\ket{v_1, \downarrow}}\xspace}

\newcommand{\evs}[0]{\ensuremath{E_{\mathrm{vs}}}\xspace}
\newcommand{\devs}[0]{\ensuremath{\Delta E_{\mathrm{vs}}}\xspace}
\newcommand{\gp}[0]{\ensuremath{g_{\mathrm{push}}}\xspace}
\newcommand{\fres}[0]{\ensuremath{f_{\mathrm{res}}}\xspace}
\newcommand{\fR}[0]{\ensuremath{f_R}\xspace}
\newcommand{\bext}[0]{\ensuremath{B_{\mathrm{ext}}}\xspace}
\newcommand{\tmw}[0]{\ensuremath{t_{\mathrm{mw}}}\xspace}
\newcommand{\fmw}[0]{\ensuremath{f_{\mathrm{mw}}}\xspace}
\newcommand{\pt}[0]{\ensuremath{P_{\mathrm{T}}-\langle P_{\mathrm{T}}\rangle}\xspace}
\newcommand{\tts}[0]{\ensuremath{T_2^*}\xspace}

\newcommand{\figlabel}[1]{\text{#1}}
\DeclareSIUnit \electronvolt { eV } 
\DeclareSIUnit\permille{\text{\textperthousand}}
\begin{document}
\title{Enhanced intrinsic spin-orbit driving of a Loss-DiVincenzo qubit near the spin-valley hotspot in Si/SiGe}
\author{Alexander Willmes}
\thanks{These two authors contributed equally}
\affiliation{JARA-Institute for Quantum Information, RWTH Aachen University and Forschungszentrum Jülich GmbH, D-52074 Aachen, Germany}
\author{Max Oberländer}
\thanks{These two authors contributed equally}
\affiliation{JARA-Institute for Quantum Information, RWTH Aachen University and Forschungszentrum Jülich GmbH, D-52074 Aachen, Germany}
\author{Max Beer}
\affiliation{JARA-Institute for Quantum Information, RWTH Aachen University and Forschungszentrum Jülich GmbH, D-52074 Aachen, Germany}
\author{Denny Dütz}
\affiliation{JARA-Institute for Quantum Information, RWTH Aachen University and Forschungszentrum Jülich GmbH, D-52074 Aachen, Germany}
\author{Jhih-Sian Tu}
\affiliation{Helmholtz Nano Facility (HNF), Forschungszentrum Jülich GmbH, D-52428 Jülich, Germany}
\author{Stefan Trellenkamp}
\affiliation{Helmholtz Nano Facility (HNF), Forschungszentrum Jülich GmbH, D-52428 Jülich, Germany}
\author{Marco Lisker}
\affiliation{IHP - Leibniz Institute for High Performance Microelectronics, D-15236 Frankfurt (Oder), Germany}
\author{Felix Reichmann}
\affiliation{IHP - Leibniz Institute for High Performance Microelectronics, D-15236 Frankfurt (Oder), Germany}
\author{Lars R. Schreiber}
\affiliation{JARA-Institute for Quantum Information, RWTH Aachen University and Forschungszentrum Jülich GmbH, D-52074 Aachen, Germany}
\affiliation{ARQUE Systems GmbH, D-52074 Aachen, Germany}
\author{Hendrik Bluhm}
\altaffiliation{bluhm@physik.rwth-aachen.de}
\affiliation{JARA-Institute for Quantum Information, RWTH Aachen University and Forschungszentrum Jülich GmbH, D-52074 Aachen, Germany}
\affiliation{ARQUE Systems GmbH, D-52074 Aachen, Germany}
\begin{abstract}
In most Si/SiGe-based spin qubit implementations, high-fidelity single-qubit gates are achieved using micromagnets, which enable the use of electric spin dipole resonance via synthetic spin-orbit coupling (s-SOC).
In contrast, intrinsic spin-orbit coupling (i-SOC) in silicon is generally considered to be weak. 
However, in Si/SiGe heterostructures, theory predicts a substantial enhancement when the Zeeman splitting approaches the valley splitting if symmetry is reduced by an imperfect interface.
Here, we demonstrate a Si/SiGe Loss-DiVincenzo qubit driven by i-SOC close to this so-called spin-valley hotspot.
In particular, we characterize the Rabi frequency as a function of the energy detuning from the hotspot by sweeping both the magnetic field and quantum dot position.
We observe the predicted enhancement of the Rabi frequency near the hotspot, but also find an asymmetry that deviates from existing theoretical models as well as distortions of the Chevron patterns near the hotspot.
While we achieve an average single-qubit Clifford fidelity of $\SI{98.6}{\percent}$, the strong variability of the valley splitting may impede the use of i-SOC-based control as a scalable operational strategy; understanding its effect is nevertheless important for reproducible high-fidelity control.
Our results provide an empirical basis for refining current theoretical models of spin-valley physics in Si/SiGe heterostructures.
\end{abstract}
	
\maketitle
\section{Introduction}
Si/SiGe spin qubits have recently shown single-qubit gate fidelities $>99.999\%$ \cite{takedaAssessingFidelitylimiting2026,wuSimultaneousHighFidelity2026}, two-qubit gate fidelities sufficient for error correction \cite{xueQuantumLogic2022,noiriFastUniversal2022,millsTwoqubitSilicon2022}, and short readout times \cite{takedaRapidSingleshot2024}.
Moreover, scalability challenges are being actively addressed, such as industrial fabrication \cite{membersofthehrlquantumteamandcollaboratorsDigitallyControlled2026,george12SpinQubitArrays2025}, long-range qubit interconnects \cite{beerConveyormodeElectron2026,desmetHighfidelitySinglespin2025,struckSpinEPRpairSeparation2024,dijkemaCavitymediatedISWAP2025} and cryogenic control electronics \cite{veldhorstSiliconCMOS2017, barteeSpinqubitControl2025, thomasRapidCryogenic2025, membersofthehrlquantumteamandcollaboratorsDigitallyControlled2026}.\par
Most current implementations of Loss-DiVincenzo (LD) qubits in Si/SiGe employ micromagnets to utilize synthetic spin-orbit coupling (s-SOC), allowing for electric spin dipole resonance (EDSR) as a means of driving qubit rotations \cite{pioro-ladriereElectricallyDriven2008}.
This commonly provides reasonably deterministic addressability and a relatively homogeneous drive strength across an array of qubits \cite{wuSimultaneousHighFidelity2026}.
In contrast, the intrinsic spin-orbit coupling (i-SOC) in Si is often assumed to be weak \cite{burkardSemiconductorSpin2023,pradaSpinOrbit2011,nestoklonSpinValleyorbit2006}.
However, reduced symmetry due to imperfect quantum well interfaces in a heterostructure may significantly enhance i-SOC strength \cite{hosseinkhaniTheorySilicon2022,jacobsonAnisotropicSpinvalley2026}.
Moreover, the valley splitting plays a significant role: A substantial enhancement of spin-orbit coupling is predicted near the so-called \emph{hotspot} \cite{huangFastSpinvalleybased2021,burkardSemiconductorSpin2023}.
Such hotspots occur when the valley splitting (\evs) equals the Zeeman splitting ($E_B$) supplied by an external magnetic field (\bext), allowing the spin and valley states to hybridize.
However, hotspots are also associated with enhanced spin relaxation and decoherence \cite{huangFastSpinvalleybased2021,burkardSemiconductorSpin2023}.\par
\begin{figure*}
    \centering
    \includegraphics[width=\textwidth]{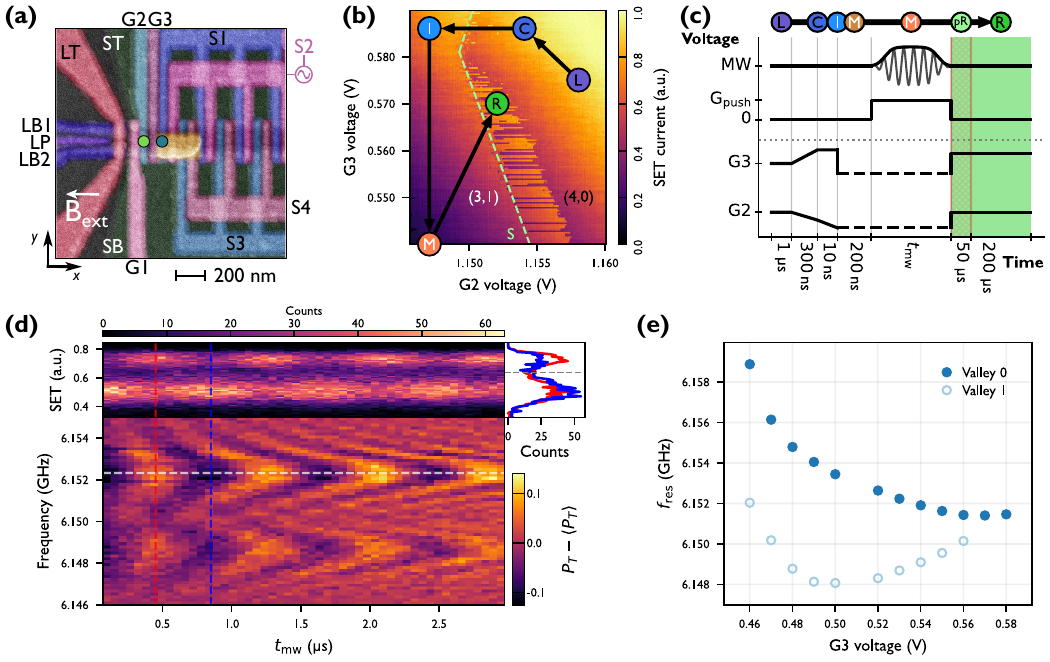}
    \caption{
        \label{fig:exposition}
        \figlabel{(a)} False-colored scanning electron microscopy image of a nominally identical device where the green (blue) dot indicates the position of the left (right) quantum dot.
        A cobalt nanomagnet (amber rectangle) was evaporated on top of the gate stack but was found not to produce a detectable magnetic field (see main text).
        A microwave source is connected to gate set S2 for driving EDSR in the right dot.
        An external magnetic field \bext is applied in-plane along the channel.
        \figlabel{(b)} Charge stability diagram of the closed system in the (3,1)-(4,0) regime, obtained by pulsing the interdot barrier and the plunger gate of the first dot; (L,R) denotes electron occupancies of left and right dots.
        The Pauli spin blockade (PSB) region is visible at the interdot charge transition.
        Circles indicate pulse positions for loading (L), crossing (C), initialization (I), manipulation (M), and readout (R).
        \figlabel{(c)} Rabi experiment pulse sequence.
        \figlabel{(d)} Rabi oscillations.
        (Bottom): Dual-valley occupancy Chevron pattern as a function of microwave frequency and burst duration.
        (Top): SET signal counts as a function of microwave duration at $\fmw= \SI{6.1523}{\giga\hertz}$ (white dashed line). 
        Oscillations between the blocked (high signal) and non-blocked (low signal) state are superimposed on a drive-independent background.
        The top-right inset shows line cuts through the SET signal at burst durations corresponding to maximum blocked (red) and non-blocked (blue) probability.
        The gray dashed line indicates the singlet/triplet discrimination threshold.
        \figlabel{(e)} Qubit resonance frequencies $\fres$ of the two valleys shown in (d) vary as a function of interdot barrier voltage G3 due to changes in position and confinement.
    }
\end{figure*}
On a theoretical level, operation based on intrinsic spin-orbit coupling has been proposed in Refs. \cite{woodsSpinorbitEnhancement2023,soomroHighfidelityEDSR2026}, in conjunction with heterostructure modifications.
Experiments have shown i-SOC-driven singlet-triplet oscillations \cite{caiCoherentSpin2023} in a Si/SiGe double quantum dot and incoherent spectroscopy signatures in fully depleted silicon on insulator (FDSOI) nanowires \cite{cornaElectricallyDriven2018}.
Here, we implement a Loss-DiVincenzo qubit in an isotopically purified (800 ppm) $^{28}$Si/SiGe heterostructure and show coherent control near the spin-valley hotspot using i-SOC (\cref{sec:experiment}).
Starting in \cref{sec:fR(Evs)}, we investigate the magnitude of the spin-valley coupling by probing the resonance frequency (\fres) as a function of \bext and find values an order of magnitude larger than previously observed in Si/SiGe devices.
From Rabi oscillations in the vicinity of the hotspot, we find the expected enhancement of the Rabi frequency (\fR), but observe an asymmetry of the peak not captured by the theoretical model of Ref. \cite{huangFastSpinvalleybased2021}.
Additionally, we see the quality of Chevron patterns deteriorate near the hotspot, showing that the hotspot cannot be straightforwardly exploited for high-fidelity control.
Lastly, to show basic operability of a qubit driven by i-SOC, we perform calibration and randomized benchmarking experiments, obtaining an average single-qubit Clifford fidelity of $F\approx\SI{98.6}{\percent}$, limited by noise on the resonance frequency (\cref{sec:iSOCQubit}) of ultimately inconclusive origin, although we speculate that nuclear spins may play a role.
The use of i-SOC as the primary drive mechanism in larger architectures may be impracticable as the valley splitting, and consequently the hotspot for fixed $E_B$, is subject to significant fluctuations due to alloy disorder \cite{paqueletwuetzAtomicFluctuations2022}.
Nevertheless, i-SOC can enable the operation of individual qubits without micromagnets and may be useful for characterization, e.g. of regions of shuttling channels far from any micromagnet. A thorough understanding of i-SOC is desirable for assessing the effect of its interplay with micromagnet-mediated EDSR, for example regarding qubit variability and control fidelity.
\section{Experiment}
\label{sec:experiment}
\Cref{fig:exposition}(a) shows a scanning electron micrograph of a nominally identical device.
The device, which is similar to the one used in Ref. \cite{xueSiSiGe2024}, consists of a shuttling section (gates S1--S4), a static quantum dot (gates G1--G3), and a single-electron transistor (SET).
In the present experiment, we form a double dot within the static dot and the beginning of the shuttling section, as indicated by the green and blue dots in \cref{fig:exposition}(a).
The right (blue) dot can be moved within the channel using the shuttling gates.
In order to manipulate the spin state of the blue qubit we make use of electric dipole spin resonance (EDSR) and apply microwave pulses on gate S2, in addition to baseband arbitrary-waveform-generator (AWG) pulses.
A cobalt nanomagnet, evaporated over the beginning of the shuttling section, was intended to provide a magnetic-field gradient for EDSR driving.
To magnetize the nanomagnet as well as induce a Zeeman splitting, we apply an external magnetic field \bext along the channel as indicated in \cref{fig:exposition}(a).
However, we found no evidence for any magnetic field produced by the nanomagnet (see \cref{sec:ext:no_uMagnet}) even after ramping the external field to $\SI{2}{\tesla}$.
Post-experiment magnetic force microscopy did not reveal a detectable magnetic signature, while energy-dispersive X-ray spectroscopy measurements suggested that the cobalt nanomagnet may have oxidized to an antiferromagnetic cobalt oxide.
\cref{fig:exposition} panels (b) and (c) describe the basic operating procedure of our experiments. 
We operate the system in the (4,0)-(3,1) charge regime to enhance the width of the Pauli spin blockade (PSB) region \cite{philipsUniversalControl2022}.
After loading four electrons into the left dot as a singlet state, we close the tunnel barrier to the reservoir by pulsing G1 to a high negative voltage relative to the other gates (L).
To facilitate a fast interdot charge transition, the voltage on G3 is increased to lower the interdot tunnel barrier (C).
A rapid adiabatic passage to the (3,1) charge regime is performed by pulsing on G2, intending to traverse the singlet-triplet anticrossing diabatically (I) but adiabatically map the singlet state onto the $\ket{\uparrow,\downarrow}$ spin configuration.
The electron in the right dot, whose spin serves as the qubit, is then isolated (M) and moved along the channel.
Although the device in principle enables charge movement over large distances using the shuttling gates S1-S4, we were unable to attain qubit manipulation in the device tuning regimes necessary for shuttling the electron.
Hence, we confine ourselves to the first double quantum dot (DQD) and move the electron only by a few nanometers using a virtual gate $\gp$, with $(\Delta\mathrm{G3},\Delta\mathrm{S2})=\gp \cdot(-1, 2.5)$ (defined relative to an initial operating point for both gates).
At these positions, the qubit is subjected to microwave driving via gate S2 for a duration of \tmw.
Finally, the system is pulsed to the readout position (R) where the parity of the double-dot spin configuration is measured via PSB for $\SI{200}{\micro\second}$ using room temperature lock-in detection at $\SI{37.3}{\kilo\hertz}$.
The readout is preceded by a $\SI{50}{\micro\second}$ wait to allow the filter of the lock-in to settle.
For all spin experiments, we repeat the sequence many times and calculate the average-subtracted triplet probability \pt by fitting two Gaussians to the histogram of the SET current and thresholding at the crossover point (\cref{fig:exposition}(d) top right panel).
In our experiments, we typically obtain a visibility of $\SIrange{20}{30}{\percent}$ with a drive-independent and fluctuating background.
The exact origin of this background remains unclear.
We speculate that intervalley tunneling and a weak Zeeman gradient contribute to overall state-preparation-and-measurement errors.
To compensate for the slow fluctuations in the measurement contrast, we perform the thresholding separately for adequately chunked data.\par
Using this procedure, we perform Rabi experiments in which we vary the microwave drive frequency and burst duration \tmw, and observe a superposition of two Chevron patterns with closely aligned Rabi frequencies (\cref{fig:exposition}(d)). 
We attribute the two patterns to the electron being initialized in either of the two valley states, which differ in visibility due to unequal initialization probabilities.
The two valley states generally have a small difference in their g-factors, typically on the order of a few per mille \cite{volmerMappingGfactors2026,kawakamiElectricalControl2014,ferdousValleyDependent2018,volmerMappingValley2024,volmerImpactLocal2026}.
As a result, the two valleys have different resonance frequencies, since $\fres=g\mu_{\mathrm{B}}\bext/h$.
Indeed, we observe a voltage-dependent difference in resonance frequency between the two valleys when varying the voltage applied to G3 during driving at point M.
This voltage variation modifies the position and confinement of the dot, thereby changing the local environment probed by the electron wave functions and impacting the $g$-factor.
The resulting difference in $g$-factor is in the range of $\SIrange[range-units = brackets, range-phrase = -]{0}{1}{\permille}$, consistent with previously reported spatial variations of the $g$-factor difference \cite{jacobsonAnisotropicSpinvalley2026,volmerMappingGfactors2026}.
In most measurements, the higher-frequency resonance (labeled Valley 0 in \cref{fig:exposition}(e)) is dominant, and we therefore focus our subsequent experiments on this transition.
\section{Rabi frequency as a function of valley splitting}
\label{sec:fR(Evs)}
\begin{figure*}
    \centering
    \includegraphics[width=\textwidth]{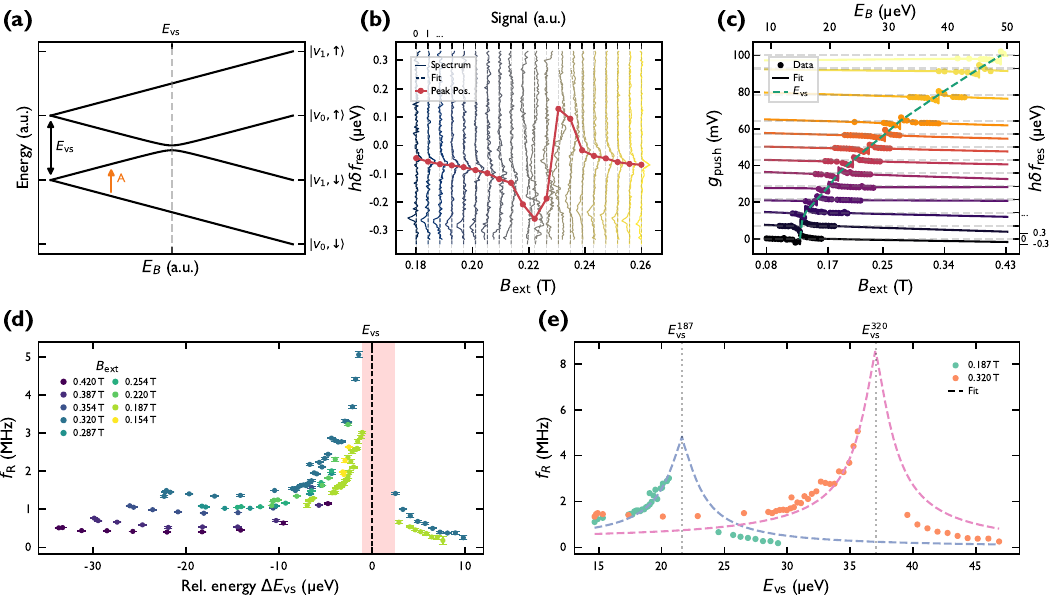}
    \caption{
        \label{fig:rabi}
        \figlabel{(a)} Energy level diagram of an electron with two valley states split by \evs as a function of Zeeman splitting $E_B$.
        The spin-valley hotspot is located at $E_B=\evs$.
        Arrow \emph{A} indicates the transition driven when the qubit is initialized in the \ld state. 
        \figlabel{(b)} Qubit spectra and resonance extraction for various external magnetic fields at $\gp=\SI{50}{\milli\volt}$.
        Each trace (top axis) shows the normalized SET signal as the microwave frequency is swept around the expected resonance, $h\delta\fres=h\fres-g\mu_B\bext$ (left axis), at fixed burst time $t_{\mathrm{mw}}$.
        A second (negative) peak at lower frequencies results from the subtraction of a frequency-detuned reference pulse.
        The extracted peak positions from Gaussian fits (ruby red) exhibit an anticrossing near $\SI{230}{\milli\tesla}$, as expected for a spin-valley hotspot.
        \figlabel{(c)} Anticrossing positions for various \gp. For each \gp, the gray dashed line and inner ticks on the right correspond to the zero position of $\delta \fres$; the outer ticks indicate the scale.
        Fitting \cref{eq:EVSanticross} to the anticrossing traces yields \evs and $\abs{C}$ and maps \evs as a function of \gp (green dashed line). 
        \figlabel{(d)} Rabi frequency \fR vs. distance from the spin-valley hotspot $\devs=\evs-E_B$, controlled via \gp, at several external magnetic fields.
        \fR increases drastically near the hotspot, and has asymmetric flanks.
        In the region marked in red, no reliable extraction of \fR is possible.
        \figlabel{(e)} i-SOC model fit (\cref{eq:SOCmodel}) to $\fR(\evs)$ at $\bext=\SI{187}{\milli\tesla}$ and $\bext=\SI{320}{\milli\tesla}$ using an absolute energy axis. The model captures the enhancement near the hotspot but not the strong asymmetry from the high energy tail; possible explanations are discussed in the main text.
    }
\end{figure*}
Our main result is the measured dependence of the Rabi frequency on energy detuning from the spin-valley hotspot, achieved by altering dot position and magnetic field.
As a first step, we map out the valley splitting by probing the resonance frequency along the virtual gate axis \gp using spectroscopy pulses.
The expected behavior of the resonance frequency can be understood from \cref{fig:rabi}(a), which shows the energy diagram of a single spin with two valley states separated by \evs as a function of Zeeman splitting $E_B$. 
When $E_B$ is not close to \evs, the resonance frequency for spin transitions within the same valley satisfies $h\fres=E_B=g\mu_\mathrm{B}\bext$. 
Crucially, at $E_B=\evs$, the states \lu and \ud undergo an anticrossing where spin and valley states hybridize, known as the spin-valley hotspot, and where this frequency relation deviates. 
For a fixed position (\gp-value), stepping \bext and subtracting a linear slope from extracted resonance frequencies yields a characteristic anticrossing pattern (\cref{fig:rabi}(b)), with the crossing center being the spin-valley hotspot.
This readily provides both the magnitude of the valley splitting as a function of \gp (\cref{fig:rabi}(c)) as well as the magnitude of the spin-valley coupling matrix element $C/2$, which determines the width of the anticrossing.
From a basic Hamiltonian
\begin{equation}
    H=\frac{1}{2}\left(\evs\tau_z+E_B\sigma_z+C\tau_x\otimes\sigma_x\right)\,,
\end{equation}
where $\tau_i$ and $\sigma_i$ are the Pauli matrices and $E_\pm=\evs\pm E_B$, the relevant resonance frequencies are
\begin{equation}
    \label{eq:EVSanticross}
    \fres^{\mathrm{fit}}=\frac{1}{2h}
    \begin{cases}
        \sqrt{E_+^2+C^2}-\sqrt{E_-^2+C^2}\,,\,\, E_B<\evs\\
        \sqrt{E_+^2+C^2}+\sqrt{E_-^2+C^2}\,,\,\, E_B>\evs\\
    \end{cases}\,.%
\end{equation}
Subtracting the linear slope of the Zeeman contribution gives $\delta\fres=\fres-g\mu_B\bext/h$.
We find $\abs{C}\approx\SI{0.75}{\micro\electronvolt}$, approximately equal for all analyzed \gp positions.
This value is larger than previously extracted spin-valley couplings from measurements of singlet-triplet oscillations in Si/SiGe by an order of magnitude \cite{volmerMappingValley2024,jacobsonAnisotropicSpinvalley2026,caiCoherentSpin2023}.
However, comparable spin-valley coupling strengths have been reported in SiMOS and FDSOI nanowire devices \cite{jacobsonAnisotropicSpinvalley2026,cornaElectricallyDriven2018}.
In Si/SiGe, a similar enhancement may arise from quantum well interface roughness \cite{hosseinkhaniRelaxationSingleelectron2021,huangElectricallyDriven2017} or from non-negligible or fluctuating Ge concentration in the quantum well \cite{woodsSpinorbitEnhancement2023}.
\par
We go on to perform measurements of the Rabi frequency in proximity to this hotspot by recording Chevron patterns at the resonance frequencies previously determined (\cref{sec:ext:fr_extract}).
The resulting dependence of \fR on the detuning from the hotspot, $\devs=\evs-E_B$, is depicted in \cref{fig:rabi}(d).
Three immediate observations can be made:
Firstly, the Rabi frequency is significantly enhanced near the hotspot, as is expected from theory \cite{huangFastSpinvalleybased2021, cornaElectricallyDriven2018}.
Secondly, this enhancement is asymmetric around the hotspot, settling toward a roughly constant nonzero value for $\evs<E_B$, but rapidly approaching zero for $\evs>E_B$.
As discussed later, the observed degree of asymmetry is not captured by a model presented in Ref. \cite{huangFastSpinvalleybased2021}.
Thirdly, the quality of the recorded Chevron patterns significantly deteriorates close to the hotspot, prohibiting reliable extraction of \fR.
This is marked as a red region in panel (d) and discussed in \cref{sec:ext:fr_extract}.
\par
We further record data for different magnetic fields.
For all fields where reliable measurements are possible, the data show the same trend as described above.
The hotspot condition $\evs=E_B$ implies different $\evs$ and thus different positions of the dot for these measurements according to our $\evs(\gp)$-relation.
The common trend therefore underlines that the observed effect mainly depends on the proximity of \evs to the hotspot in energy space, and is not the result of further valley-independent atomistic details.
While we try to account for frequency-dependent attenuation effects of our microwave power and aim for a constant external drive strength at the sample, this calibration may be associated with an uncertainty of order unity when comparing significantly different frequencies.
The difference in \fR at comparable detuning even in the saturated region for different magnetic fields may hence arise from calibration uncertainties.
\par
An enhancement of \fR around the spin-valley hotspot has been predicted in literature \cite{huangFastSpinvalleybased2021,cornaElectricallyDriven2018}.
Huang and Hu \cite{huangFastSpinvalleybased2021} model spin-valley effects in the four-dimensional spin-valley-space.
Near the hotspot, mixing between the $\ket{\uparrow,v_0}$ and $\ket{\downarrow,v_1}$ states becomes relevant, where $v_i$ denotes the valley state and $\uparrow/\downarrow$ denote spin states.
As Huang and Hu calculate, \fR can be written as
\begin{equation}
    \label{eq:SOCmodel}
    \fR\sim\abs{\bra{v_0}\hat{x}\ket{v_1}}\cdot
    \begin{cases}
        \abs{\sin\left(\frac{\gamma_+ \pm\gamma_-}{2}\right)}\,,\quad E_B<\evs\\
        \abs{\cos\left(\frac{\gamma_+ \pm\gamma_-}{2}\right)}\,,\quad E_B>\evs\\
    \end{cases}%
    \,,
\end{equation}
where the definition $\gamma_\pm = \text{atan2}(|C|,E_\pm)$ is used and $\bra{v_0}\hat{x}\ket{v_1}$ is the dipole matrix element between both valley states.
The dipole matrix element is treated as a phenomenological parameter as it depends sensitively on the microscopic details of the interface.
Huang and Hu show that two spin-valley mixing paths contribute to the driving, and that their relative phase depends on the nature of SOC: s-SOC breaks time-reversal symmetry while i-SOC does not, leading to constructive or destructive interference of the two paths, respectively.
\cref{eq:SOCmodel} shows the case of pure synthetic SOC ('$+$'-path) and intrinsic SOC ('$-$'-path) for brevity.
The data in \cref{fig:rabi}(e), however, are fitted using the complete formula from Ref. \cite{huangFastSpinvalleybased2021} which includes mixing of both spin-valley paths.
The mixing parameter naturally converges toward the pure-i-SOC case, which increases the asymmetry between $E_B<\evs$ and $E_B>\evs$ compared to s-SOC, but as $\gamma_+\ll\gamma_-$ in our regime this difference is small regardless.
As can be seen from the fit, the model only qualitatively captures the data and exhibits increasing deviations farther away from the hotspot and for $E_B>\evs$.
This indicates that \cref{eq:SOCmodel} cannot be the full picture, or the intervalley dipole element $\bra{v_0}\hat{x}\ket{v_1}$ depends nontrivially on the dot position.
The latter is not unreasonable to assume, as our position variation with \gp likely also changes the wavefunction and the local alloy disorder that it probes.
Assuming in the simplest case a linearized variation, fit quality for both datasets individually is, expectedly, notably increased.
However, we are unable to obtain consistent global, i.e., magnetic-field-independent parameters for such a dependency.
It appears that the asymmetry arises relative to the hotspot rather than from a dependence purely on the dot position.\par
\section{i-SOC qubit characterization}
\subsection{Gate fidelity}
\label{sec:iSOCQubit}
\begin{figure}
    \centering
    \includegraphics[width=\columnwidth]{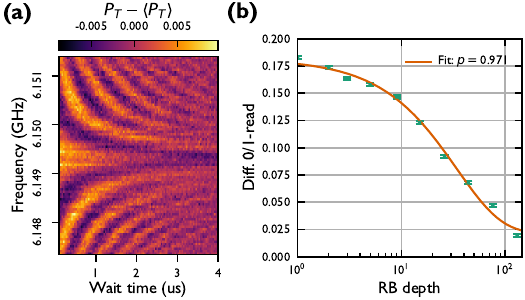}
    \caption{
    Qubit characterization at $\bext=\SI{220}{\milli\tesla}$, tuned away from the hotspot with $E_B>\evs$.
    \figlabel{(a)} Ramsey experiment at the same electrostatic configuration used for randomized benchmarking in (b).
    Fitting decaying oscillations (\cref{eq:decayingoscillation}) to the Ramsey pattern yields a median $T_2^*=\SI{2.3}{\micro\second}$.
    In general, the measured $T_2^*$-values are strongly position-dependent (see \cref{sec:ext:coherenceVsgp}).
    \figlabel{(b)} Randomized benchmarking (RB) following AllXY calibration.
    \label{fig:coherence}}
\end{figure}
To demonstrate elementary qubit operation, we benchmark the i-SOC-driven spin qubit using Clifford randomized benchmarking (RB) \cite{magesanScalableRobust2011}.
To this end, we first calibrate the single-qubit gate set using the AllXY gate sequence.
The resulting error syndromes of this 21-circuit protocol allow us to identify drive errors such as amplitude, detuning, and skew errors \cite{reedEntanglementQuantum2013}.
We then carry out the RB experiment (utilizing \texttt{pyGSTi} \cite{nielsenProbingQuantum2020}) with a native gate set of $X_{\pi/2}$, $Y_{\pi/2}$ (and identity operation) at an operating point of $\bext=\SI{220}{\milli\tesla}$ with a Rabi frequency of $\fR\approx\SI{0.85}{\mega\hertz}$.
\cref{fig:coherence}(b) shows the RB decay curve at the optimal microwave power, yielding a decay parameter $p=0.971$ and thus an average single-qubit Clifford fidelity of $F_{\mathrm{avg}}=\SI{98.6}{\percent}$.
The corresponding infidelity is approximately three orders of magnitude larger than the state-of-the-art reported for spin qubits driven via s-SOC with micromagnets \cite{wuSimultaneousHighFidelity2026}.
While further optimization of, e.g., \fR, which is based here on only rudimentary parameter space exploration, or the application of strategies like dedicated pulse shaping \cite{wuSimultaneousHighFidelity2026} would likely improve the fidelity, we expect that the significant frequency noise, manifested as a short \tts-time as discussed below, ultimately prohibits this quantity from approaching state-of-the-art values.
Our data further show that the quality factor decreases monotonically when approaching the hotspot (see \cref{sec:ext:coherenceVsgp}), indicating that no improvement in fidelity from operating closer to the hotspot is to be expected.
\subsection{Qubit coherence (\texorpdfstring{$T_2^*$}{T2s})}
Ramsey measurements at the same electrostatic operating point as the fidelity measurement, shown in \cref{fig:coherence}(a) with each trace recorded for about \SI{2}{\minute} of laboratory time, yield $T_2^*\approx\SI{2.3}{\micro\second}$. 
In conjunction with the Rabi frequency of $\SI{0.85}{\mega\hertz}$, this number is broadly consistent with the Clifford fidelity obtained from RB although it is significantly shorter than several values reported elsewhere for the same isotopic purity \cite{yonedaQuantumdotSpin2018,neyensProbingSingle2024}.
However, reported values fluctuate significantly even between nominally identical quantum dots \cite{philipsUniversalControl2022,millsTwoqubitSilicon2022,stanoReviewPerformance2022}.
In isotopically purified Si, the $T_2^*$-time is commonly thought to be limited by charge noise that couples to the qubit through the magnetic field gradient of the micromagnet, as shown in the case of Ref.\,\cite{yonedaQuantumdotSpin2018}.
In the present device, this error channel is absent due to the nonfunctional on-chip magnet, but the proximity to the hotspot provides an alternative mechanism through which charge noise can couple to the qubit. 
It is thus interesting whether an analysis of the coherence of this sample can shed new light on decoherence mechanism in Si/SiGe spin qubits.
Measurements of charge noise in the device as well as other samples from the same fabrication run reveal a relatively large value of $\sqrt{S(\SI{1}{\hertz})}\approx\SI{24}{\micro\electronvolt\per\sqrt\hertz}$ with a typical $1/f$-slope (see \cref{sec:ext:gatenoise}).
While the high charge noise level is near a plausible order of magnitude to account for the observed dephasing, the measured dependence of $T_2^*$ on the energy detuning from the hotspot is weaker than predicted by a model of the avoided crossing if incorporating solely gate-equivalent charge noise.
A better match is obtained when including an additional uncorrelated noise contribution with a fitted rms-magnitude of \SI{140}{\kilo\hertz}, which is about a factor five too large to be explained with a measured background dependence of the qubit frequency on gate voltage (\cref{fig:ext:coherenceVsgp}).
Moreover, the Chevron patterns near the hotspot tend to be highly distorted in a way that is not easily accounted for by charge noise (\cref{fig:ext:rabiweirdexamples}).
For \devs-values several microelectronvolts below zero, the Chevrons appear regular, as expected (\cref{fig:ext:rabiexamples}(a)).
As the hotspot is approached, the Chevron patterns become irregular and sometimes exhibit abrupt jumps or discontinuities (\cref{fig:ext:rabiweirdexamples}).
Very close to the hotspot (red shaded region in \cref{fig:ext:rabiexamples}(h,i)), oscillations remain visible, but the characteristic Chevron pattern disappears (\cref{fig:ext:rabiexamples}(g)).
Beyond the hotspot ($\devs\gg0$) the Chevrons qualitatively recover their regular appearance, although distortions persist.
This behavior is quantified by the phenomenological $k$-factor used in the Rabi frequency extraction (\cref{eq:Rabi} and \cref{fig:ext:rabiexamples}(i)), which decreases from $1$ to about $0.45$ as \devs crosses zero.
Qualitatively, it is consistent with drifts or abrupt jumps of the resonance frequency by several $\SI{}{\mega\hertz}$ during measurement.
More generally, the Chevrons extend over a much broader drive frequency range than expected for the observed Rabi frequency, as reflected by the reduced $k$-factor.\par
We speculate that residual silicon and germanium nuclear spins could contribute both additional frequency noise and nontrivial dynamics close to the hotspot.
In particular, nuclear-spin-mediated feedback between the microwave drive and the qubit resonance, similar to mechanisms reported previously \cite{vinkLockingElectron2009,lairdHyperfineMediatedGateDriven2007}, could cause the resonance frequency to follow the drive frequency before abruptly switching back, thereby producing distorted Chevron patterns.
Such interaction effects may be strongly enhanced by proximity to the hotspot, which makes energetically nearby electronic states with opposite spin available. 
In conjunction with the enhanced spin-orbit coupling, they can thus enable dynamic nuclear polarization (DNP) with repeated angular momentum transfer. 
Likewise, electron-mediated nuclear flip-flop transitions may enhance spin diffusion in the nuclear system, which would speed up sampling of an ergodic ensemble of nuclear configurations and could thus lead to shortest measured values of \tts.
DNP effects involving positive feedback, contrary to the case considered in Refs \cite{tenbergNarrowingOverhauser2015,bluhmEnhancingCoherence2010}, may even enhance the rms frequency noise due to nuclear noise compared to the commonly assumed value of order $A/\sqrt{N}$ obtained for statistically independent random nuclear spin polarization, where $A$ is the hyperfine coupling constant and $N$ the number of significantly coupled nuclear spins.
Consistent with this hypothesis, one of the longest coherence times in Si/SiGe reported to date was measured in a qubit with a higher degree of purification of 60 ppm \cite{struckLowfrequencySpin2020}, whereas most other data were obtained at a residual concentration of 800 ppm.
\section{Conclusion}
\label{sec:conclusion}
In conclusion, we demonstrated a Si/SiGe LD qubit driven by intrinsic spin-orbit coupling without (efficacious) nanomagnet or on-chip electron-spin-resonance line, and investigated the dependence of its Rabi frequency on the energy detuning from the spin-valley hotspot at different magnetic fields.
Interestingly, we observe a rather strong intrinsic spin-orbit coupling, with a coupling matrix element about one order of magnitude larger than previously reported in this material composition \cite{volmerMappingValley2024,jacobsonAnisotropicSpinvalley2026,caiCoherentSpin2023}.
Consistent with theoretical expectations, we observe an enhancement of the Rabi frequency near the spin-valley hotspot.
However, the Rabi frequency exhibits an asymmetry that is not captured by the simple four-state model considered here and requires further theoretical investigation.
Recent theoretical work \cite{soomroHighfidelityEDSR2026} has derived a valley-splitting-dependent term for the Rabi frequency, but has not yet considered operation close to the hotspot as studied here.\par
Randomized benchmarking yielded an average single-qubit Clifford fidelity of $\SI{98.6}{\percent}$, which remains significantly below the state-of-the-art for EDSR-driven spin qubits.
In addition, the measured $T_2^*$ is shorter than expected for isotopically purified silicon, and the observed deformed Chevron patterns point to a significant noise source whose origin ultimately remains unresolved.
Although charge noise may contribute, it seems questionable as a sole cause and we find anecdotal indications that nuclear spin effects are enhanced near the hotspot. 
This would suggest that nuclear spins may play a more significant role for qubit coherence in isotopically purified silicon than is commonly appreciated and that particular care needs to be taken near spin-valley hotspots.
While higher fidelities should be achievable using i-SOC, relying on energetic proximity to the hotspot for operating multi-qubit devices faces the challenge that the magnetic field at which the hotspot occurs varies along with the valley splitting between different quantum dots.
However, being able to manipulate qubits using i-SOC may be a useful tool for characterizing device regions that do not have any micromagnet, such as shuttling channels, and isolating different decoherence mechanisms.
i-SOC-related corrections to Rabi driving primarily related to s-SOC may also be relevant for understanding qubit variability.
Understanding the magnitude of i-SOC as well as the noise enhancement and distortion of Rabi-Chevrons requires further investigation.
\section*{Acknowledgments}
We thank Marcus Liebmann for conducting MFM analysis on the device.\par
This work was funded by the German Research Foundation (DFG) within the project 421769186 (SCHR 1404/5-2) and under Germany's Excellence Strategy - Cluster of Excellence Matter and Light for Quantum Computing (ML4Q) EXC 2004/2 - 390534769.
Part of this work was supported by the German Federal Ministry of Research, Technology and Space within the joint project QUASAR.
Part of this work has been carried out within the Joint Lab “Spin-Based Quantum Computing” established between IHP - Leibniz Institute for High Performance Microelectronics, RWTH Aachen University and  Forschungszentrum Jülich.
Sample fabrication was carried out at the Helmholtz Nano Facility (HNF) at the Forschungszentrum J\"ulich \cite{albrechtHNFHelmholtz2017}.
\\
\section*{Data availability}
The data that support the findings of this article are openly available \cite{willmesDataEnhanced2026}.
\\
\section*{Author contributions}
M.L. and F.R. prepared the heterostructure and fabricated parts of the device used in this study (markers, ohmic contacts).
M.B., D.D., J.S.T. and S.T. carried out remaining fabrication. A.W. and M.O. set up and conducted the experiment, analyzed the data and prepared the manuscript with input from all authors.
L.R.S. and H.B. provided guidance in data interpretation and supervised the project.
\section*{Competing interests}
L.R.S. and H.B. are founders and shareholders of ARQUE Systems GmbH. The other authors declare no competing interests.
\appendix
\setcounter{section}{0}
\renewcommand{\thesection}{\Alph{section}}
\crefalias{section}{appendix}

\makeatletter
\renewcommand{\section}[1]{%
    \refstepcounter{section}%
    \par\bigskip%
    \begin{center}%
        \textbf{Appendix \thesection: #1}%
    \end{center}%
    \par\bigskip%
}
\makeatother

\counterwithin{figure}{section}
\counterwithin{table}{section}
\setcounter{figure}{0}
\setcounter{table}{0}

\section{Experimental details}
\label{sec:ext:expdetails}
The device was fabricated on a commercial $^{28}$Si/SiGe heterostructure epitaxially grown by chemical vapor deposition.
The qubit is defined in a tensile-strained $^{28}$Si quantum well with a thickness of $\SI{7}{\nano\meter}$ and with 800\,ppm residual $^{29}$Si, located beneath a barrier layer of $\SI{30}{\nano\meter}$ $^{28}$Si$_{0.7}$Ge$_{0.3}$ and capped by a layer of $\SI{1}{\nano\meter}$ $^{28}$Si.
The quantum well was grown on a strain-relaxed \SI[quantity-product = {-}]{1000}{\nano\meter}-thick Si$_{0.7}$Ge$_{0.3}$ layer, which was grown on a chemically mechanically polished virtual substrate. The virtual substrate consists of a \SI[quantity-product = {-}]{1000}{\nano\meter}-thick relaxed Si$_{0.7}$Ge$_{0.3}$ layer above a \SI[quantity-product = {-}]{3000}{\nano\meter}-thick linearly graded buffer layer grown on a Si(100) substrate.
Ohmic contacts were defined by selective phosphorus ion implantation with a dose of \SI{4.5e15}{\per\square\centi\metre} at an implantation energy of \SI{20}{\kilo\electronvolt}.
The implanted wafers were annealed for \SI{1}{\minute} at \SI{700}{\celsius} to activate the phosphorus dopants.
Overlapping Ti-Pt gate electrodes were fabricated by \SI{100}{\kilo\electronvolt} electron-beam lithography, metal evaporation and lift-off, with $\SI{10}{\nano\meter}$ of Al$_2$O$_3$ as an inter-gate dielectric. The main shuttling axis is aligned along the (110)-crystallographic axis.\par
As a step toward simplifying wiring requirements, we omit bias tees on most of our gates, except for G1 and the screening gates.
We DC-couple either Zurich Instruments HDAWG or Harvard DecaDAC outputs to the remaining barrier and plunger gates, depending on the respective gate's role in our operation.
We work around the constraints of limited voltage ranges of the AWG outputs and heat dissipation at attenuators by globally offsetting all ohmic contacts of the device, which are connected to a common DAC-channel, by $\SI{-700}{\milli\volt}$ relative to system ground.
Confinement is restored by applying large negative offsets to the surrounding screening gates, as well as to G1 to obtain a closed tunnel barrier from the DQD to the SET.
Readout of charge occupation in the channel via the SET, we AC-couple a $\sim\SI{40}{\kilo\hertz}$ excitation of amplitude $\SI{150}{\micro\volt}$ to the ohmic contacts with a bias tee, and measure the resulting current with a room-temperature Basel Precision Instruments SP983c01-IF transimpedance amplifier subsequently feeding into a Zurich Instruments MFLI lock-in amplifier.
\section{Rabi frequency extraction}
\label{sec:ext:fr_extract}
\begin{figure*}[!t]
    \centering
    \includegraphics[width=\textwidth]{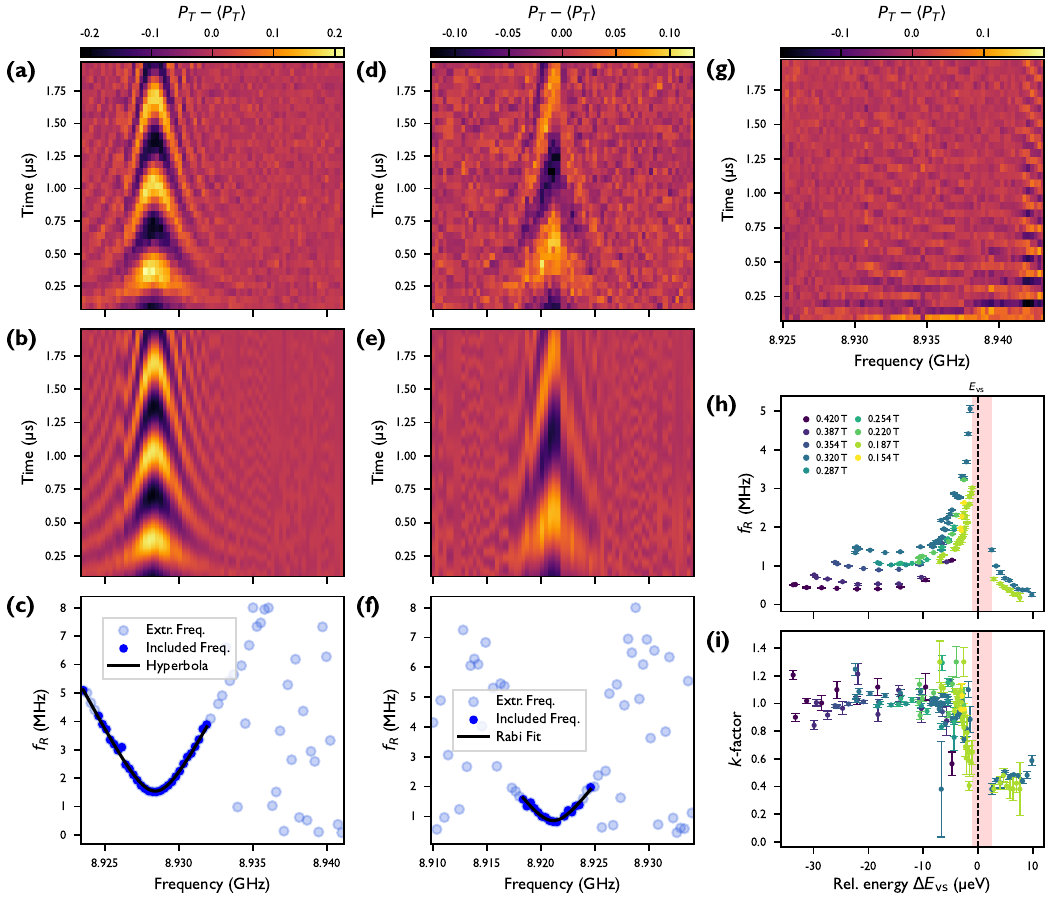}
    \caption{Rabi frequency extraction workflow at $\bext=\SI{320}{\milli\tesla}$ for different positions.
    \figlabel{(a,d,g)} Raw data obtained for $\evs<E_B$ ($\devs=\SI{-8.2}{\micro\electronvolt}$), $\evs>E_B$ ($\devs=\SI{4.1}{\micro\electronvolt}$), and $\evs\sim E_B$ ($\devs=\SI{-1.1}{\micro\electronvolt}$), respectively. For (g), frequency extraction was omitted due to insufficient data quality. Both valley resonance frequencies are expected within the measured frequency range.
    \figlabel{(b,e)} Fits of \cref{eq:decayingoscillation} to the data in (a) and (d) at each microwave drive frequency. 
    \figlabel{(c,f)} Extracted oscillation frequencies from (b) and (d). A fit of \cref{eq:Rabi} (black line) yields the on-resonance Rabi frequency $f_R$. A set of filtering parameters selects data points that are included (dark blue) and excluded (light blue) to ensure robust fitting. Data in (d)-(f) require a phenomenological factor $k$ to accurately match the width of the Chevron. (h) Extracted Rabi frequencies for various external magnetic fields and \gp values (same as \cref{fig:rabi}). (i) Corresponding $k$-factors from \cref{eq:Rabi}. Notably, $k\approx1$ for $\devs\ll0$, while $k\approx0.45$ when close to or above the hotspot for $\devs\gtrsim 0$. 
    \label{fig:ext:rabiexamples}}
\end{figure*}
\begin{figure}[!t]
    \centering
    \includegraphics[width=\columnwidth]{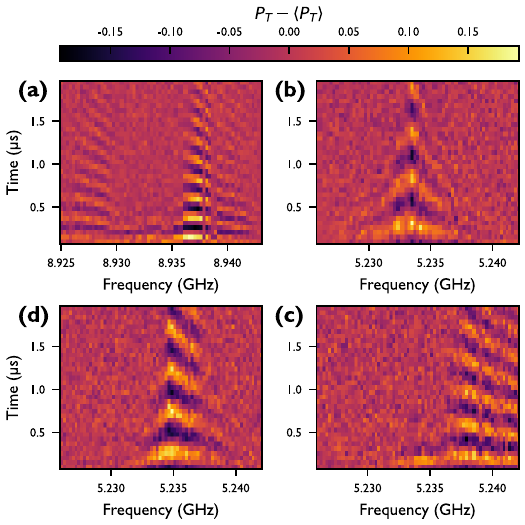}
    \caption{Examples of distorted Chevron patterns near the spin-valley hotspot for
    \figlabel{(a)} $\devs=\SI{-1.8}{\micro\electronvolt}$ and $\bext=\SI{320}{\milli\tesla}$, 
    \figlabel{(b)} $\devs=\SI{-2.9}{\micro\electronvolt}$ and $\bext=\SI{187}{\milli\tesla}$, 
    \figlabel{(c)} $\devs=\SI{-2.0}{\micro\electronvolt}$ and $\bext=\SI{187}{\milli\tesla}$, 
    \figlabel{(d)} $\devs=\SI{-1.6}{\micro\electronvolt}$ and $\bext=\SI{187}{\milli\tesla}$. 
    The reappearance of the oscillation in (a) can be attributed to driving in the other valley.
    \label{fig:ext:rabiweirdexamples}}
\end{figure}

To extract Rabi frequencies from Chevron patterns, we use a two-step procedure.
First, for each microwave frequency separately, we fit the qubit oscillation frequency using an exponentially decaying sinusoid
\begin{equation}
\label{eq:decayingoscillation}
Ae^{-(\frac{\tmw}{\tau})^\kappa}\sin{(2\pi f \tmw + \phi)}+c
\end{equation}
where $A$ is the oscillation amplitude, $\tau$ is the decay time constant, $\kappa$ is the decay exponent, $f$ is the oscillation frequency, $\phi$ is the phase, and $c$ is a readout offset.
In a second step, we assign the extracted frequencies $f(\fmw)$ to the two valleys and fit each branch separately with the Rabi formula
\begin{equation}
    \label{eq:Rabi}
    f=\sqrt{\fR^2+k(\fmw-\fres)^2}\,.
\end{equation}
Here, we intentionally introduce a phenomenological scaling factor $k$ to the (squared) frequency detuning to be able to fit our data.
Data points for the second fitting stage are pre-selected by applying range-based thresholds to the extracted $A$, $f$ and $\tau$ from \cref{eq:decayingoscillation}.
The resulting selection is manually reviewed to correct cases where data points are incorrectly included or excluded, for example when points from the wrong valley are selected.
\cref{fig:ext:rabiexamples} illustrates the Rabi frequency extraction from the Chevron pattern for three different cases: one where $E_B>\evs$ and $k\approx1$ (a-c), one where $E_B<\evs$ and $k<1$ (d-f), and one at $E_B \approx \evs$ where the Rabi frequency extraction is not possible (g).
Away from the hotspot on the $\devs<0$ side, the phenomenological factor $k$ in \cref{eq:Rabi} remains close to $1$.
Notably however, when approaching the hotspot and when $\devs>0$, $k$ exhibits substantial spread and decreases to values as low as $k\approx0.45$.
This variation in $k$ is accompanied by distortions of the Chevron patterns near the hotspot.
In particular, the patterns become irregular and exhibit abrupt jumps and sudden cut-offs, examples of which are shown in \cref{fig:ext:rabiweirdexamples}.
In general, values of $k<1$ indicate that the Rabi frequency changes more weakly with microwave frequency than expected.
One possible explanation could be a repulsive interaction between the resonance and drive frequency due to an additional physical effect, as has been observed from nuclear spins during sweeps of an external magnetic field \cite{vinkLockingElectron2009}.
In this picture, the interaction shifts the resonance ahead of the probed frequency until an abrupt jump occurs, potentially explaining both the sudden jumps as well as values of $k<1$.
At the same time, the reduction in $k$ may also be an artifact caused by abrupt jumps in the resonance frequency.
Regardless of the underlying origin, the extracted Rabi frequencies for $k<1$ cannot be the cause for an increased effective asymmetry of \fR around the peak:
Even abrupt resonance-frequency jumps are expected, in the worst case, to lead only to an artificial increase of the fitted Rabi frequency. 
However, the observed asymmetry of the Rabi frequency with respect to the hotspot as seen in \cref{fig:rabi}(e) is characterized by lower \fR for $\devs>0$, precisely where $k<1$.
\section{Nanomagnet characterization}
\label{sec:ext:no_uMagnet}
\begin{figure}[!t]
    \centering
    \includegraphics[width=\columnwidth]{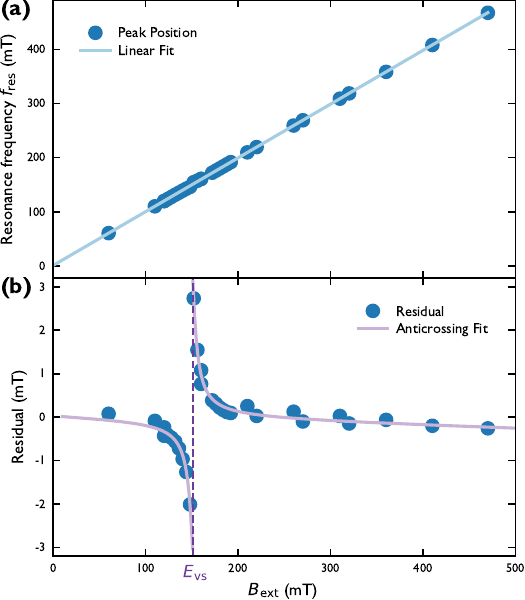}
    \caption{
    \figlabel{(a)} Extracted resonance frequency from qubit spectroscopy given in units of \si{\milli\tesla} as a function of external magnetic field. A linear fit of the form $m\bext+b$ yields $m=\SI{0.995}{}$ and $b=\SI{0.70}{\milli\tesla}$. $m=1$ and $b=0$ corresponds to the case where the sole field contribution stems from the external magnet. The fit, therefore, indicates a negligible contribution from the nanomagnet and consequently to the observed driving. \figlabel{(b)} Residuals of linear fit from (a). Fit to \cref{eq:EVSanticross} yields the position of the spin-valley hotspot at $\bext=\SI{151}{\milli\tesla}$ corresponding to $\evs=\SI{17.5}{\micro\electronvolt}$ and a coupling strength of $C=\SI{0.74}{\micro\electronvolt}$.
    \label{fig:ext:noMagnet}}
\end{figure}

As mentioned in the main text, we find no evidence that the nanomagnet provides a magnetic field of detectable magnitude.
We base this conclusion on two observations:
First, tracking the resonance frequency as a function of \bext and fitting a linear slope to the data (see \cref{fig:ext:noMagnet}(a)) yields a residual magnetization of only \SI{0.7}{\milli\tesla}, far below the contribution of $\geq \SI{10}{\milli\tesla}$ expected from the geometry of the nanomagnet.
Even though previous empirical assessments of the stray field of such magnets suggest deviations from idealized simulations \cite{aldeghiSimulationMeasurement2025,philipsUniversalControl2022}, potentially due to the polycrystalline structure of the cobalt used, we do not expect such deviations to obscure a functional nanomagnet with hysteretic magnetization.
To further verify this we employed magnetic force microscopy analysis of the nanomagnet at room temperature after completion of the experiment, which likewise returned no indication of magnetization.
Subsequent energy-dispersive X-ray spectroscopy indicated a significant peak of oxygen at the position of the nanomagnet, a possible signature of oxidation, but this evidence is not conclusive on its own because the finite EDX spot size may include residual influence from surrounding oxide.
\section{$T_2^*$-\gp dependence}
\label{sec:ext:coherenceVsgp}
\begin{figure}[!t]
    \centering
    \includegraphics[width=\columnwidth]{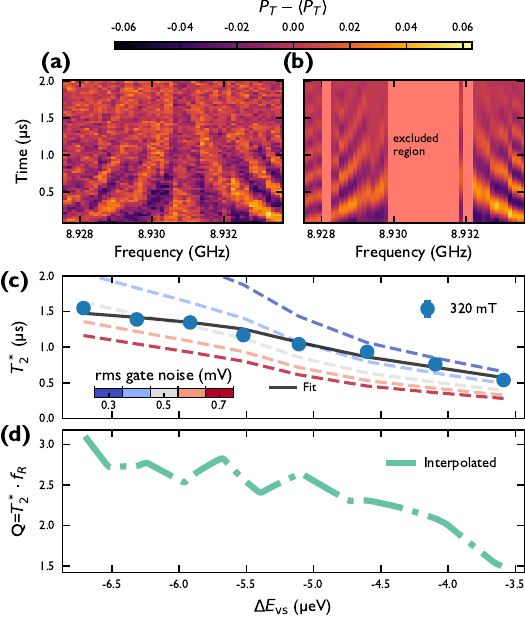}
    \caption{
    \figlabel{(a)} Ramsey measurement performed at \mbox{$\bext=\SI{320}{\milli\tesla}$} and $\devs=\SI{-5.5}{\micro\electronvolt}$.
    \figlabel{(b)} Fit of \cref{eq:Ramsey} to the data shown in (a).
    The light red region marks fits excluded in post-selection based on minimal visibility and frequency detuning.
    \figlabel{(c)} Extracted $T_2^*$ as a function of \devs (points), comparison to expected slopes from effective transferred gate noise (dashed lines), and fit with baseline term and effective gate-referred charge noise rms as free parameters (solid line).
    \figlabel{(d)} Quality factor as a function of \devs, declining toward the hotspot.
    \label{fig:ext:coherenceVsgp}}
\end{figure}
We record Ramsey patterns while approaching the hotspot, which we fit with
\begin{equation}
    \label{eq:Ramsey}
    P_T=s\cdot\exp\left[-\left(t/\tau\right)^k\right]\sin\left(2\pi\delta \fres t+\phi\right)\,,
\end{equation}
as displayed in \cref{fig:ext:coherenceVsgp}(a, b).
Here, $k$ is fixed to $2$ for consistency with later heuristic analytical treatment.
To ensure fit-reliability, we retain only fits post-selected based on a required minimal visibility of the oscillation and a minimum detuning from the resonance.
We compare extracted $\tts(\devs)$ values to a predicted slope derived from the susceptibility of \fres to fluctuations of $\Delta\evs$ at the operation point, implicitly assuming that gate-referred variations of \evs are a good proxy for all charge effects in the vicinity.
Using \cref{eq:EVSanticross} to get the derivative of \fres with respect to the valley splitting, one obtains for $\evs<E_B$
\begin{equation}
    \label{eq:fResByEvs}
    \pdv{\fres}{\gp}=\frac{1}{2h}\pdv{\evs}{\gp}\left(\frac{E_-}{\sqrt{E_-^2+C^2}}+\frac{E_+}{\sqrt{E_+^2+C^2}}\right)\,,
\end{equation}
and $\pdv{\evs}{\gp}$ can be retrieved from the data shown in \cref{fig:rabi}(c).
Additional, uncorrelated frequency fluctuations may be absorbed into the quantity $\sigma_f^{\mathrm{base}}$, such that the total frequency variation reads
\begin{equation}
    \label{eq:fResByG}
    \sigma_f=\sqrt{\left(\sigma_{\gp}\pdv{\fres}{\gp}\right)^2+(\sigma_f^{\mathrm{base}})^2}\,.
\end{equation}
In a first step, we neglect $\sigma_f^{\mathrm{base}}$ and plot \mbox{$\tts=1/(\sqrt{2}\pi\pdv{\fres}{\gp}\sigma_{\gp})$} (treating $\sigma_{\gp}$ as the effective equivalent Gaussian quasistatic rms value within the given measurement protocol, which is consistent between measurement points) in \cref{fig:ext:coherenceVsgp} next to experimental data as a function of energy detuning \devs from the hotspot, assuming different exemplary rms-values for effective gate-referred noise $\sigma_{\gp}$.
Although, expectedly, both measured $T_2^*$-times as well as analytic slope decrease when approaching the spin-valley hotspot, the measured data exhibits a weaker dependence on $\devs$ compared to the analytic expectation when assuming purely this enhanced sensitivity to \evs-fluctuations (and by extension, charge fluctuations) as the cause for the decline in coherence time.
Allowing now a finite baseline $\sigma_f^{\mathrm{base}}$ and fitting to the data yields $\sigma_{\gp}=\SI{0.32}{\milli\volt}$ and $\sigma_f^{\mathrm{base}}=\SI{140}{\kilo\hertz}$.
The origin of this baseline term remains speculative.
\cref{sec:ext:gatenoise} shows that the susceptibility of frequency shifts with respect to gate voltages away from the hotspot is much weaker than above $\sigma_f^{\mathrm{base}}$ would require, amounting with fitted $\sigma_{\gp}$, when conservatively rounding $\pdv{\fres}{g_{\mathrm{far}}}\approx\SI{100}{\kilo\hertz\per\milli\volt}$, to only $\tilde{\sigma}_f=\SI{32}{\kilo\hertz}$, assuring that there is likely no elevated and wrongfully disregarded baseline of this charge-noise susceptibility.
One may consider nuclear spin noise as a speculative candidate to account for $\sigma_f^{\mathrm{base}}$.\par
In \cref{fig:ext:coherenceVsgp}(d), we show the quality factor \mbox{$Q=\fR\cdot T_2^*$} as function of \devs, using linearly interpolated data from previous measurements of both quantities.
$Q$ decreases approximately monotonically toward the hotspot, with no clear evidence for a maximum.
The data start at the onset of increase in \fR when approaching the hotspot (see \cref{fig:rabi}), which means it would cover any increase in $Q$ that would be caused by favorable interplay of enhanced Rabi frequency and decoherence mechanisms.
Hence, we do not observe the predicted \enquote{sweet spot} from Ref. \cite{huangFastSpinvalleybased2021}.
\section{Gate noise susceptibility}
\label{sec:ext:gatenoise}
\begin{figure}[!t]
    \centering
    \includegraphics[width=\columnwidth]{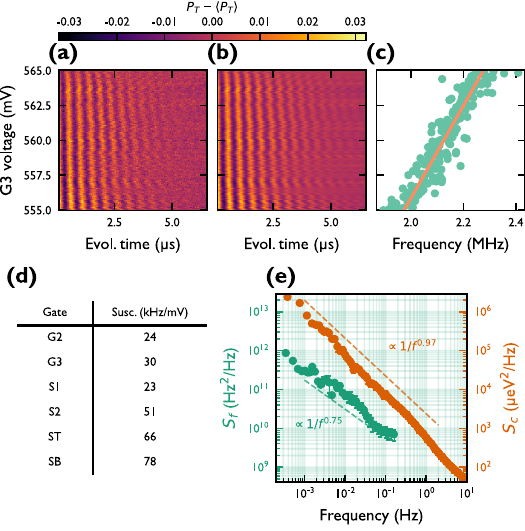}
    \caption{Frequency response of the qubit to voltages applied to nearby gates.
    \figlabel{(a)} Raw data from a single-line Ramsey measurement detuned by $\SI{1}{\mega\hertz}$ as a function of G3 voltage. 
    \figlabel{(b)} Fit of \cref{eq:Ramsey} to (a). 
    \figlabel{(c)} Extracted oscillation frequency as a function of the applied gate voltage, together with a linear fit.
    \figlabel{(d)} Frequency susceptibility of resonance frequency with respect to different gates.
    \figlabel{(e)} Spectra of the resonance frequency during a repeated AllXY experiment and of the SET current during idle, shown as green and orange curves, respectively.
    \label{fig:ext:gatenoise}}
\end{figure}
A possible origin of the lower-than-expected $T_2^*$-times and distorted Chevron patterns is charge noise or noise on the gates surrounding the qubit.
The latter is not expected to be of relevant magnitude in our setup compared to the former, but having no definitive knowledge of the absolute lever arms on the qubit, it makes sense to subsume and refer to both in terms of effective gate voltage fluctuations.
Hence, while we can estimate that with the SET-extracted charge noise spectrum plotted in \cref{fig:ext:gatenoise}(e) and typical lever arms on the order of \SI{0.1}{\electronvolt\per\volt} a magnitude necessary to explain trajectories in \cref{fig:ext:coherenceVsgp}(c) is obtained, the actual purpose of this analysis lies in aiding the discussion of \cref{sec:ext:coherenceVsgp} focusing on the slope of $\tts(\devs)$.
\par
To this end, we extract the susceptibility of the resonance frequency to each gate by performing Ramsey scanlines as a function of gate voltage at a fixed drive frequency detuned by $\SI{1}{\mega\hertz}$ from the resonance (\cref{fig:ext:gatenoise}(a)).
For each scanline, we fit \cref{eq:Ramsey} (\cref{fig:ext:gatenoise}(b)) to extract the oscillation frequency as a function of voltage (\cref{fig:ext:gatenoise}(c)).
Since the applied voltage changes are small, the corresponding frequency shifts are small, such that a linear fit is sufficient to determine the susceptibility of the resonance frequency to each gate.
The extracted values for all gates are summarized in \cref{fig:ext:gatenoise}(d).
Although the susceptibilities are measured away from the hotspot, they still contain a residual dependence on the hotspot, $\pdv{\evs}{G_i}$.
We cannot factor out this contribution, since we do not have a map for all $\evs(G_i)$.
Nevertheless, the extracted susceptibilities constitute an upper bound on any gate-noise baseline term that could dominate over the $\pdv{\evs}{\gp}$ contribution and thus serve as a conservative estimate in the analysis of \cref{sec:ext:coherenceVsgp}.
\par
\cref{fig:ext:gatenoise}(e) shows the spectra of the resonance frequency during a repeated AllXY experiment over approximately $9.5$ hours (green) and of the SET current of a nominally identical device during idle on the flank of a Coulomb oscillation (orange).
The latter was measured over a larger frequency range than the data from the device used for the rest of the study; however, both spectra overlap in their common frequency region.
Notably, the slopes do not fully align, with the SET-extracted charge noise exhibiting a typical $1/f^\alpha$ spectrum with $\alpha\approx1$, while the resonance frequency spectrum -- considering the whole measured range -- is better fit by $\alpha\approx 0.75$, although exhibiting variations in its slope over the given frequency range.
This may potentially be reminiscent of nuclear spin diffusion models as in Ref. \cite{reillyMeasurementTemporal2008}, but quantitative comparison is not fruitful based on present data; Ref. \cite{rojas-ariasOriginsNoise2025} has furthermore reported unexplained deviations from this model recording data on a natural silicon heterostructure.
It is also not unreasonable to assume a frequency-dependent transfer function between charge- and resonance-frequency-noise, such that this is not a conclusive indicator for significantly contributing effects beyond the charge noise.
\bibliography{Willmes2026a}
\clearpage
\end{document}